\documentclass[twocolumn,aps,letter,preprint,10pt]{revtex4}
\usepackage{graphicx,graphics,float} 
\usepackage{dcolumn} 
\usepackage{bm} 
\usepackage{amsmath,amssymb}
\usepackage{verbatim}
\usepackage{mathdots}
\usepackage{mathrsfs}
\usepackage{hyperref}

\begin{document}
\title{Time-reversal symmetry breaking in a square lattice}
\author{Kevin Jimenez}\email{kfjimenezfals@gmail.com}
\author{Jose Reslen}\email{reslenjo@yahoo.com}
\affiliation{Coordinaci\'on de F\'{\i}sica, Universidad del Atl\'antico, Carrera 30
N\'umero 8-49,  Puerto Colombia.}%
\date{\today}

\newcommand{\Keywords}[1]{\par\noindent{\small{\em Keywords\/}: #1}}
\begin{abstract}
The bulk conductivity of a two-dimensional system is studied assuming that quantum
interference effects break time-reversal symmetry in the presence of strong
spin-orbit interaction and strong lattice potential. The study is carried out by
direct diagonalization in order to explore the nonlinear-response regime. The
system displays a quantized conductivity that depends on the intensity of the
electric field and under specific conditions the conductivity limit at zero
electric field shows a nonvanishing value.
\\
\end{abstract}
\maketitle
\section{Introduction}
\label{intro}
\begin{figure}
\begin{center}
\includegraphics[width=0.4\textwidth,angle=-0]{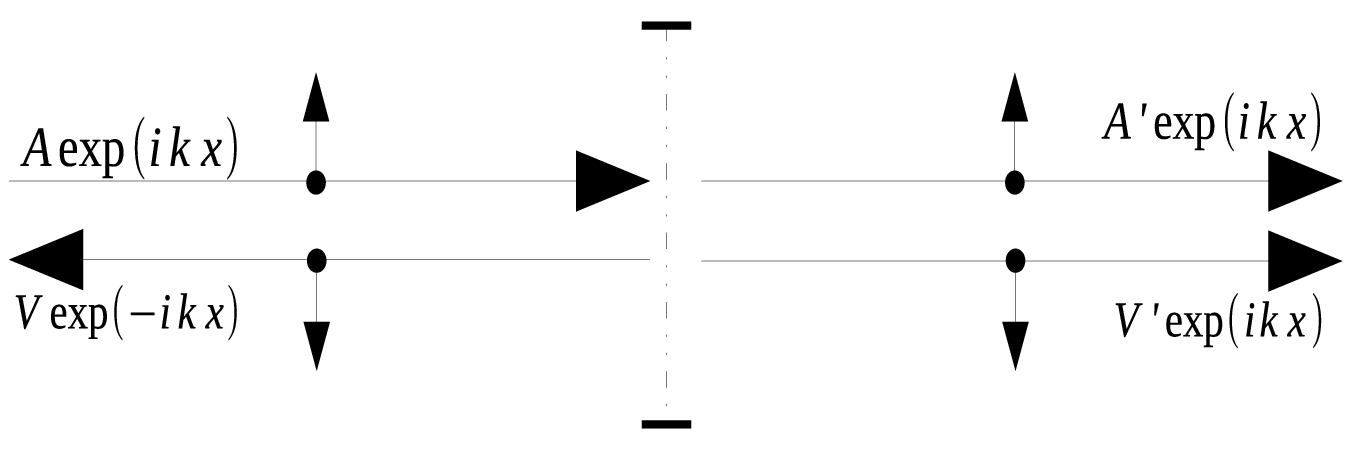} 
\caption{Scattering model of a non-magnetic obstacle.}
\label{impurity}
\end{center}
\end{figure}

An essential result from quantum mechanics prescribes that when two operators
commute there exists an eigenbasis that diagonalizes them simultaneously, so
that the elements of such an eigenbasis conform at the same time to both
operators. It is however important to highlight that this principle does not
dictate that any eigenbasis of the first operator is also an eigenbasis of the
other, which would automatically imply that any eigenstate of the first
operator should conform to the second one. Nevertheless, such an implication
takes place in one instance: When the spectrum of the first operator is
non-degenerate.  These facts are at the center of the theory of (standard)
phase transitions: In a scenario where the Hamiltonian commutes with an unitary
operator (generated by the symmetry) a critical point separates a trivial or
symmetric phase, where the state conforms to both the Hamiltonian and the
symmetry, from a non-trivial or broken phase, where the physical state is
no longer a symmetry eigenstate. A transition of this kind is only possible
when the Hamiltonian spectrum goes from non-degenerate to degenerate, being the
latter case the only one where the physical state can break the symmetry.  From
this perspective a phase transition is essentially the arising (or suppression)
of the Hamiltonian's degeneracy.  The mechanisms by which the symmetry is
broken must be on the one hand irreversible \cite{kevin}, since otherwise
equilibrium states would retain the Hamiltonian's symmetries, and on the other
hand global, so that they affect the state as a whole and the symmetry be
broken everywhere. Without these mechanisms the symmetry would not break and
the phase transition would not take place.  A typical example of this kind of
transition is the change from paramagnetic (symmetric) to ferromagnetic
(broken) in spin systems.  In the case of electron systems, a paramount result
known as the Kramers degeneracy \cite{sakurai} has significant implications in
connection to phase transitions: Time-reversal spin-systems have degenerate
spectra.  As such, the notion of phase transition in this kind of systems
cannot be accommodated in the standard symmetry-breaking paradigm associated
with a transition from degenerate to non-degenerate or vice versa.  However, it
has been observed  that a phase transition can take place whereupon the
symmetry is broken {\it locally} in the non-trivial phase, i.e., some parts of
the state, usually those associated with the system's bulk, display symmetry,
but others, like those associated with the system's boundary, do not. This
phase is known as the topological insulator\cite{zhang,zhang2,kane}. In
contrast, the trivial phase lacks symmetry-breaking mechanisms entirely, thus
being known as the standard insulator.  The symmetry associated with the
transition is {\it time-reversal} (TR) \cite{sakurai}, while the
symmetry-breaking mechanism is {\it quantum interference}, triggered
by spin-orbit interaction. In order to appreciate the mentioned mechanism with
some degree of formality, let us imagine a situation in which a spin-up
electron moving forward bumps elastically into an obstacle, as portrayed in figure
\ref{impurity}. The incident particle can be scattered in different ways but
the case is such that the only state available for backscattering corresponds
to spin down.  The obstacle effect can be modeled by the following term 
\begin{gather}
\hat{U}(x) = U (e^{i \frac{\theta}{2} \hat{\sigma}_1} + e^{-i \frac{\theta}{2}
\hat{\sigma}_1}) \delta(x), \text{ }
\hat{\sigma}_1 =
\left(
\begin{array}{cc}
0 & 1 \\
1 & 0
\end{array}
\right). 
\label{e09011}
\end{gather}
Notice that the effect is such that rotations in opposite directions by and angle
$\theta$ are equally considered. This means the potential is TR invariant, i.e., it
does not mutate under the change $t \rightarrow -t$ \footnote{This provokes $
\hat{\sigma}_1\rightarrow -\hat{\sigma}_1$}.  This case corresponds to a
non-magnetic obstacle.  The problem can be approached by considering solutions on
each side of the obstacle following the decomposition shown in figure
\ref{impurity}.
\begin{gather}
\psi_I(x) = A e^{i k x} |\uparrow \rangle + V e^{-i k x} |\downarrow \rangle,  \\
\psi_{II}(x) = A' e^{i k x} |\uparrow \rangle + V' e^{i k x} |\downarrow \rangle.  
\end{gather}
Demanding wave function continuity, $\psi_I(0) = \psi_{II}(0)$, it is found that $A
= A'$ and $V=V'$. Integration of the Schrodinger equation around the origin yields
\begin{gather}
V \left ( -\frac{ i k}{m}+ 2 U \cos \frac{\theta}{2} \right )|\downarrow \rangle  +
2 U A \cos \frac{\theta}{2} |\uparrow \rangle = 0.
\end{gather}
For finite values of $U$ the only nontrivial solution is for $V=0$ and $\theta =
\pi$, which represents a spin-up state with perfect conduction in spite of the
particle hitting an obstacle. Contrary to potential (\ref{e09011}), this perfect
conduction state is not TR invariant.  This effect is produced by the destructive
interference of backscattering paths and as such is a quantum-mechanical effect
\cite{zhang}. This result is sustained by the fact fact that backscattering is only
possible via spin inversion. The existence of spin-flipped channels in opposite
directions is guaranteed by the Kramers degeneracy as long as the Hamiltonian be TR
invariant.  The other determining factor is that the number of equal-energy
moving-channels in a given direction be odd, i.e., energy degeneracy must be an odd
multiple of two.  Contrariwise, the suppression of backscattering paths due to
quantum interference is unfeasible when the number of moving channels in one
direction is even.  This can be seen considering a Hamiltonian that displays the
common form (the constant term $\vec{\hat{\sigma}}^2$ is included only to keep a
reference to spin states) 
\begin{gather}
\hat{H}  = \frac{\vec{\hat{p}}^2}{2 m} + U(\vec{\hat{x}}) + \lambda \vec{\hat{\sigma}}^2 .
\label{josjos}
\end{gather}
It then follows $\hat{H}|-\vec{p} \rangle = \hat{H} | \vec{p} \rangle$.  As a
result, given an eigenstate with average momentum $\langle \vec{\hat{p}} \rangle$
it is always possible to construct another eigenstate with the same energy, spin
and spatial distribution, but opposite momentum $-\langle \vec{\hat{p}} \rangle$,
providing in this way direct backscattering channels. In this example the number of
moving channels on each direction is always even, because every solution admits
spin up and down. The correspondence between dissipation when there is an even
number of moving channels and conduction when there is an odd number of moving
channels can be represented using the members of the group $Z_2=\{0,1\}$. This
equivalence has prompted the use of the adjective ``topological'' when referring to
the case of nonvanishing conductivity. Also, it has been shown that it is possible
to formally establish the classification of a given system from its Block
energy-structure using topology methods \cite{z2}.  

\begin{figure}
\begin{center}
\includegraphics[width=0.4\textwidth,angle=-0]{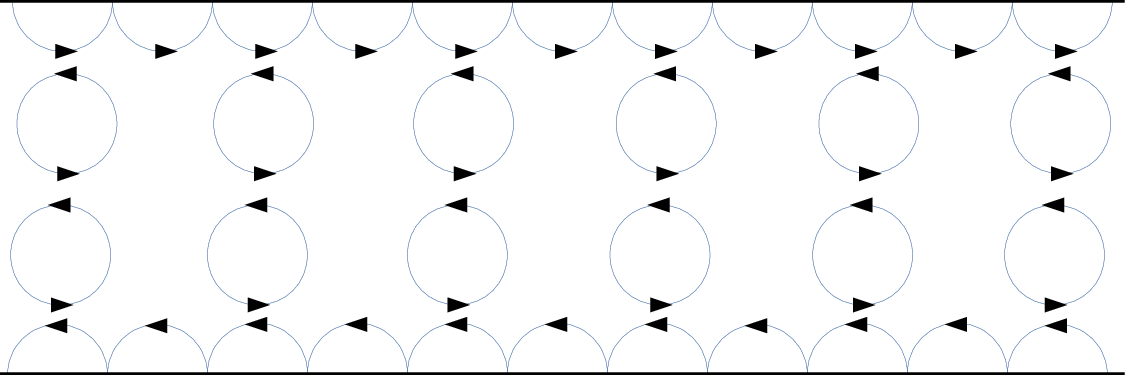} 
\caption{Boundary states break TR symmetry over the whole edge while bulk states
remain bounded.}
\label{classical_orbits}
\end{center}
\end{figure}

One way of inhibiting even numbers of moving channels is to deliberately break the
model's TR invariance.  The simplest strategy consists in applying a
magnetic field to a spinless electron 
\begin{gather}
\hat{H}  = \frac{\left( \vec{\hat{p}} - e \vec{\hat A} \right)^2}{2 m} + U(\vec{\hat{x}}).
\label{hall}
\end{gather}
In a two-dimensional space with perpendicular magnetic field the vector potential
is given by $\vec{\hat A} = (0,e B \hat{x})$. The associated term breaks the parity
symmetry since it is not invariant under the change $\vec{x} \rightarrow -\vec{x}$.
It can therefore be said that neither parity nor TR are independently preserved.
However, assuming that $U(-\vec{x}) = U(\vec{x})$, it is noticeable that the
Hamiltonian is invariant under parity and TR applied {\it simultaneously}.  In
absence of symmetry breaking mechanisms, it is at least possible to build
eigenstates that be invariant under the joined effect of both symmetries.  These
states take place both in the bulk and on the edges, as shown in figure
\ref{classical_orbits}. The interesting thing about the states that develop on the
surface is that backward-moving states shift from forward-moving states, leaving only
one moving channel on each edge and allowing the arising of an energy gap
associated with backscattering as other elements of the problem, as for example an
electric field or electron-electron collisions, are taken into account. As
backscattering channels become sufficiently suppressed, motion on the edges becomes
dissipationless.  As can be seen, the conduction mechanism is rooted in the local
breaking of the TR symmetry rather than in quantum interference, since in this
example the Hamiltonian is not TR invariant. It is known that at low temperatures
the transverse conductivity of (\ref{hall}) comes in integer multiples $n$ of
$e^2/h$ \cite{thouless,laughlin}.  This phenomenon is known as the integer quantum
Hall effect \cite{sinova}. Number $n$ has been shown to correspond to a topological
invariant \cite{kohmoto}, which explains the notable robustness of the effect
observed in experimental measurements. 
In a topological insulator dissipation channels are displaced just as in the Hall
effect, but the role of the magnetic field is taken over by the spin, such that A.
In the system's boundary the number of moving channels in one direction is odd.
And B. The Hamiltonian is TR invariant. These facts combined lead to the quantum
interference that provokes the state to break TR symmetry and so become conducting
\cite{seo,xie}.  It is often said in this regard that conduction is ``protected''
by the TR symmetry. Topological insulators are peculiar in that they can conduct
even though their bulk spectra are gaped.

The goal of the present study is to provide a numerical analysis of a single-body
model where the symmetry-breaking takes place in the bulk and is generated by a
strong lattice potential that acts in a way analogous to the edge of a topological
insulator.  It is of interest to consider the electric field as an integral part of
the problem and to observe how the conductivity depends on this field beyond the
linear approximation.  This approach intends to shed insight by helping visualize
the system's response as a complement to the more abstract analytical formulation
often found in related studies.  Interestingly, this procedure yields a quantized
conductivity that shows a dependence with the number of bands below the Fermi
energy and in some cases this conductivity remains finite as the electric field
goes to zero, suggesting in this way a superconducting state.
\section{The model and its eigensystem}
\label{l0915}
The model corresponds to an electron that moves on a two-dimensional potential
$U(x,y)$ under the action of an electric field $E$ in the $y$ direction.
Spin-orbit interaction arises as a coupling between the z-components of spin and
angular momentum. The Hamiltonian is written as
\begin{gather}
\hat{H}  = \frac{\hat{p}_x^2  + \hat{p}_y^2}{2 m} + \frac{\lambda}{m} \hat{\sigma}_z ( \hat{x} \hat{p}_y - \hat{y} \hat{p}_x) + U(\hat{x},\hat{y}) - e E \hat{y}.
\label{heart}
\end{gather}
Constants $m$ and $e$ represent mass and charge respectively. The intensity of the
spin-orbit interaction is mediated by constant $\lambda$. The potential is written
as
\begin{gather}
U(x,y) = U_x \cos \frac{2 \pi x }{a} + U_y \cos \frac{2 \pi y }{a}
\end{gather}
Current technology allows for a high degree of control over the model's parameters
in optical lattices or superlattices \cite{spielman,rey}, being these scenarios
where the effects reported further ahead are more likely to be observed.  For a
numerical analysis it is necessary to bound the system in order to provide a
compact Hilbert space, hence periodic boundary conditions are imposed on the
$x$-axis over a square lattice of side $L$.  Lattice constant $a$ is such that $L=N
a$, being $N$ the square root of the total number of real unit-cells in the
lattice. It can be noticed that due to the terms of spin-orbit and electric field,
Hamiltonian (\ref{heart}) does not display translational invariance  in neither
axis and therefore it does not admit a treatment in terms of Bloch functions.
However, it is possible to consider an alternative symmetry arising from
simultaneous translations of space and momentum, but for this it is necessary to
add a term, as follows
\begin{gather}
\hat{H}_0  = \hat{H} + \frac{\lambda^2 \hat{x}^2}{2 m}.
\label{mahe}
\end{gather}
The extra-term can be considered either as a physical confining potential, in which
case it becomes an integral part of the Hamiltonian, or as a perturbation. Both
(\ref{heart}) and (\ref{mahe}) are TR invariant, but only (\ref{mahe})
commutes with the following symmetry operator ($\hbar = 1$)
\begin{gather}
\hat{T} = e^{i a \hat{p}_x + i a \lambda \hat{\sigma}_z \hat{y}}.
\label{e07122}
\end{gather}
\begin{figure}
\begin{center}
\includegraphics[width=0.27\textwidth,angle=-90]{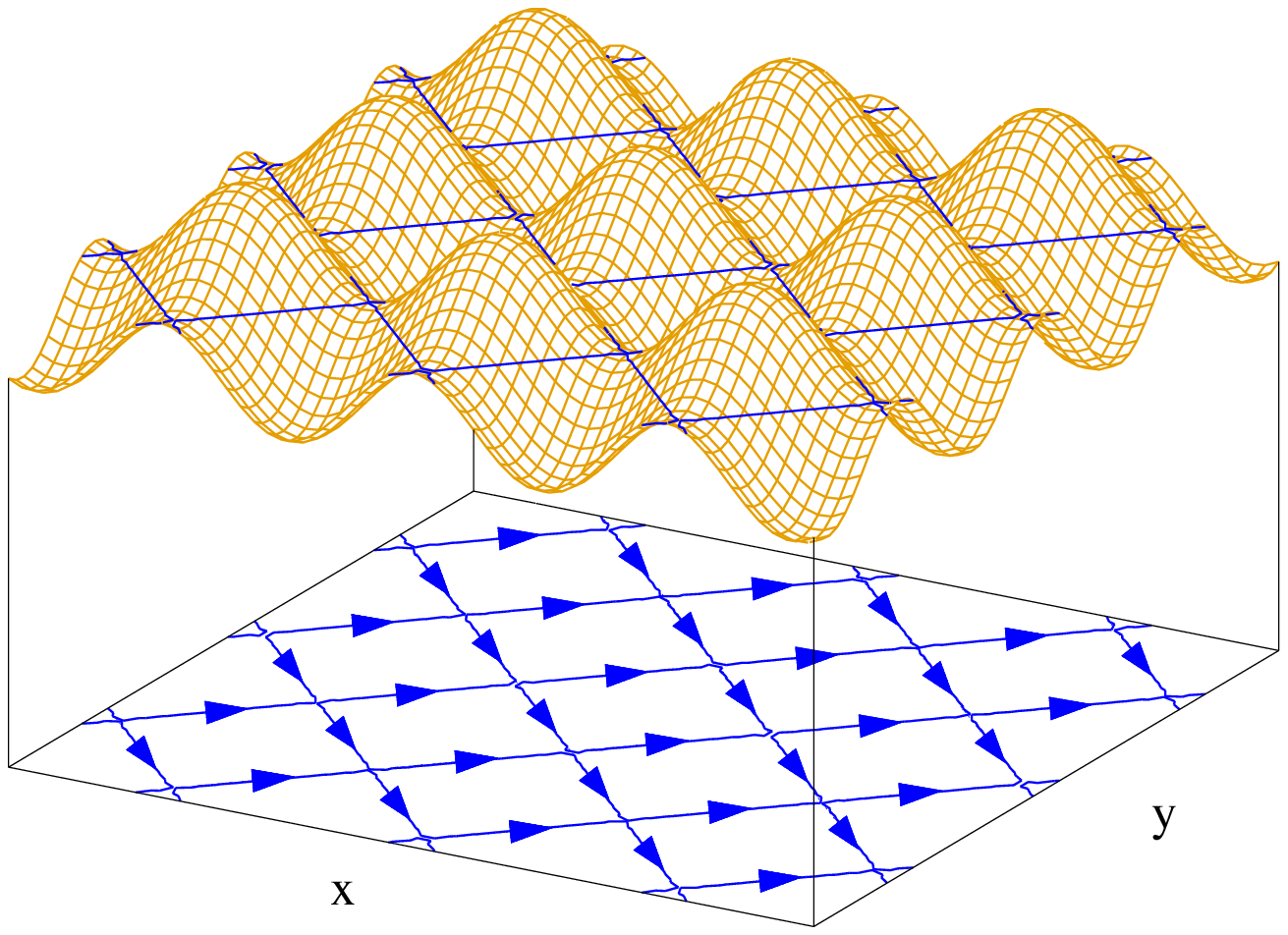} 
\caption{Conduction channels take place around potential extrema giving rise to a
net current in the $x$ axis.}    
\label{f0917}
\end{center}
\end{figure}
This can be confirmed using $\hat{T} \hat{x} \hat{T}^{-1} = \hat{x} + a$ and
$\hat{T} \hat{p}_y \hat{T}^{-1} = \hat{p}_y - a \lambda \hat{\sigma}_z$.
Fundamental results dictate that there exists a common basis for $\hat{H}_0$ and
$\hat{T}$. The most general way of writing an eigenfunction of $\hat{T}$ with
eigenvalue $e^{i k a}$ is
\begin{gather}
\psi(x,p_y,\sigma_z) = e^{i k x} u_k(x,p_y,\sigma_z),
\label{mar27}
\end{gather}
subject to the condition $u_k(x+a,p_y-a\lambda \sigma_z,\sigma_z) =
u_k(x,p_y,\sigma_z)$.  These requirements are met for functions defined as
\begin{gather}
u_k(x,p_y,\sigma_z) = \sum_{j,w} C_{j,w,\sigma_z}(k) e^{j \frac{2 \pi i}{L} x} e^{-w 2 \pi i
p_y} 
\left[
\begin{array}{c}
\delta_1^{\sigma_z} \\
\delta_{-1}^{\sigma_z} 
\end{array}
\right],
\label{botella}
\end{gather}
insofar as
\begin{gather}
\frac{j}{L} + w \lambda \sigma_z  = \frac{n}{a},
\end{gather}
being $n$ an arbitrary integer. Solving for $w$ yields
\begin{gather}
w = \frac{q}{\lambda L \sigma_z},
\label{polloloco}
\end{gather}
being $q= n N - j$. Integer $j$ is not bounded, since the corresponding momentum
eigenvalue $p_x = \frac{2 \pi j}{L}$ in (\ref{botella}) remains always consistent
with boundary conditions.  This is not the case for the position eigenvalue $y = 2
\pi w$ because the system is bounded on the $y$ axis, therefore
\begin{gather}
|2 \pi w | \le \frac{L}{2}.
\end{gather}
Using (\ref{polloloco}) it then follows
\begin{gather}
q_{max} = \frac{\lambda L^2 }{4 \pi}.
\end{gather}
Since $q$ in (\ref{polloloco}) can take negative values, the total number of
position states is given by $Q=2 q_{max} + 1$.  Inserting $Q$ and solving for $L$
gives
\begin{gather}
L = \sqrt{\frac{2 \pi (Q-1)}{\lambda}}.
\end{gather}
As a consequence, the system length depends on the number of states and the
interaction constant.  This conditioning certainty arises from the symmetry and
seems to be related to the fact that Heisenberg's uncertainty principle establishes
a phase space grating. Another result is that the system's eigenfunctions are
periodic in the $p_y$-space with period $\lambda L$, as can be seen from
(\ref{polloloco}) and (\ref{botella}).  Besides, periodic boundary conditions on
the $x$ axis $\psi(x+L,p_y)=\psi(x,p_y)$ determine as valid values of $k = \frac{2
\pi l}{L}$, for $l$ integer. The size of a unit cell in $k$-space is
$\frac{2\pi}{a}$.  The eigenvalue problem can be formulated in terms of the
symmetry functions as 
\begin{gather}
| \hat{H}_k u_k \rangle  = \mathscr{E} | u_k \rangle,
\end{gather}
wherein
\begin{gather}
\hat{H}_k = e^{-i k \hat{x}} \hat{H}_0 e^{i k \hat{x}} = \frac{(\hat{p}_x + k)^2}{2 m}  + \frac{(\hat{p}_y + \hat{\sigma}_z \lambda \hat{x})^2}{2 m} + \nonumber \\
- \frac{\lambda}{m} \hat{\sigma}_z \hat{y} (\hat{p}_x + k) + U(\hat{x},\hat{y}) - e E \hat{y},
\label{aqva}
\end{gather}
being $\mathscr{E}$ the system's energy. As the problem is separable with respect
to spin, it is valid to set $\hat{\sigma}_z=1$ in order to focus on the spin-up
case. Technically, the resulting expression breaks TR symmetry, but the operation
is justified by the fact that quantum interference breaks such a symmetry through
the same quantum interference effect that takes place in the boundary of a
topological insulator, with the difference that in this case the role of the
boundary is taken over by the periodic potential, which must in addition be
sufficiently strong to induce trajectories surrounding potential peaks in the bulk,
as sketched in figure \ref{f0917}. This mechanism must derive in an energy spectrum
whose degeneracy comes in odd multiples of two, which is to be checked further
ahead. Since $\hat{x}$ and $\hat{p}_y$ are to be used as a complete set of
commuting observables, the following results must be considered, $p_x = -i
\partial_x \text{, } y = i \partial_{p_y}$. Basis functions are taken as
eigenfunctions of $\hat{p}_x$ and $\hat{y}$ normalized over the $x p_y$-cell
\begin{gather}
\langle x,p_y | j, q \rangle = \phi_{j,q}(x,p_y)= \frac{e^{j \frac{2 \pi i}{L} x} e^{-q \frac{2 \pi i}{\lambda L} p_y}}{L \sqrt{\lambda}}.
\end{gather}
Calculation of the matrix elements, $\langle j', q'| \hat{H}_k^{\uparrow} | j, q
\rangle$, yields
\begin{figure}
\begin{center}
\includegraphics[width=0.3\textwidth,angle=-90]{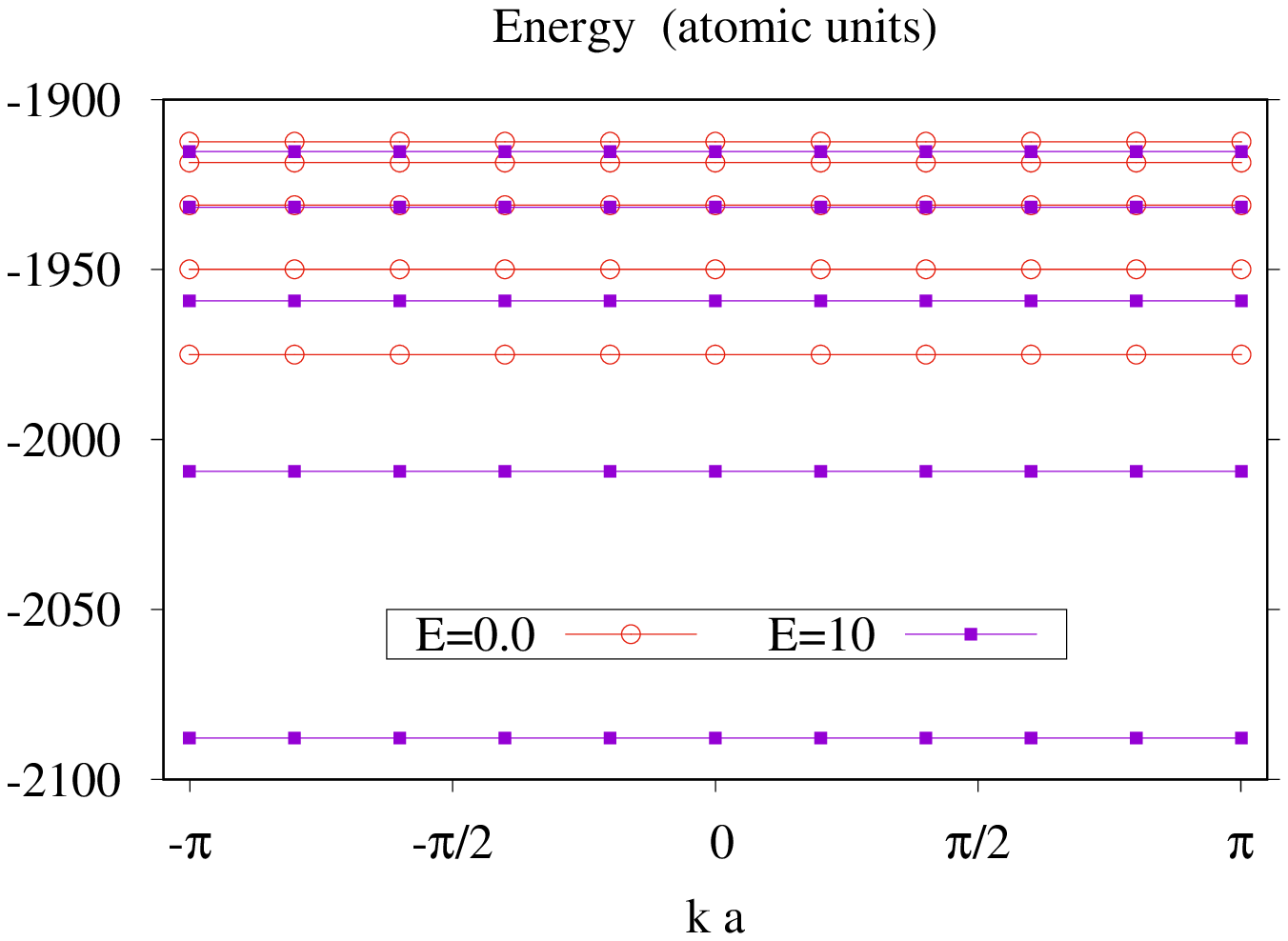} 
\caption{First bands of Hamiltonian (\ref{may23}) for extreme values of the
electric field. The model constants in atomic units are $m=1$, $e=1$, $\lambda=1$,
$U_x = U_y = 10^3$ and $L = 25.1$. Additionally  $N=10$, $Q=101$ and $J=201$.  The
band pattern is notably flat along a wide range of values of $E$. Because the
lattice potential is strong this system would be a standard insulator in absence of
spin-orbit interaction. These energies are two-fold degenerate for $E=0$ but
non-degenerate for $E=10$, being this the case where quantum interference
effects can break TR symmetry.} 
\label{f06082}
\end{center}
\end{figure}
%
%
\begin{gather}
\int_{-\frac{L}{2}}^{\frac{L}{2}} d x  \int_{-\frac{\lambda L}{2}}^{\frac{\lambda L}{2}} d p_y  \phi_{j',q'}^*(x,p_y)  H_k^{\uparrow} \phi_{j,q}(x,p_y)  = \nonumber \\
\frac{1}{2 m} \left( \frac{2 \pi j }{L} + k \right )^2 \delta_j^{j'} \delta_q^{q'} + \frac{\lambda^2 L^2}{4 \pi^2 m} \left (  f(q,q') \delta_j^{j'} + \right . \nonumber \\
\left . g(q,q') g(j,j') +  f(j,j') \delta_q^{q'}   \right ) - \frac{  2 \pi  q  }{m L}  \left( \frac{2 \pi j }{L} + k \right ) \delta_{q}^{q'} \delta_{j}^{j'} \nonumber \\
+ \frac{U_x}{2} (\delta_{N}^{j-j'} + \delta_{-N}^{j-j'}) \delta_{q}^{q'} +  U_y \cos \frac{4 \pi^2 q}{a \lambda L } \delta_j^{j'} \delta_q^{q'} + \nonumber \\
	- e E \frac{ 2 \pi  q }{\lambda L} \delta_{q}^{q'} \delta_{j}^{j'}, 
\label{may23}
\end{gather}
%
%
where
\begin{gather}
\begin{array}{cc}
	f(n,n') =
\begin{cases}
 \frac{\pi^2}{6}               & \text{ if } n = n', \\
 \frac{(-1)^{n-n'}}{(n-n')^2}  & \text{ if } n \ne n',
\end{cases}
\end{array}
\label{china}
\end{gather}
and
\begin{gather}
\begin{array}{cc}
	g(n,n') =
\begin{cases}
             0             & \text{ if } n = n', \\
 \frac{(-1)^{n-n'}}{n-n'}  & \text{ if } n \ne n'.
\end{cases}
\end{array}
\label{china}
\end{gather}
Interestingly, matricial elements are all real even though the basis functions are
complex. This helps reduce computation costs.  When this Hamiltonian is numerically
diagonalized the respective eigenfunctions take the next form
\begin{gather}
| u_k \rangle = \sum_{q=-q_{max}}^{q_{max}} \sum_{n=-n_{max}}^{n_{max}}  c_{j,q}(k) | j,q \rangle.
\label{e07121}
\end{gather}
The momentum integer is $j= N n - q$ and the total number of momentum states is $J
= 2 n_{max}+1$. Figure \ref{f06082} presents the first bands of the spin-up-part of
th Hamiltonian for a set of reasonable parameters and strong lattice potential. In
addition to being flat, the band pattern is in general nondegenerate, except for
specific values of $E$. This can be better seen on the left side of figure
\ref{f06101}, which shows energy as a function of $E$.  Given the TR symmetry of
the whole Hamiltonian together with its degeneracy profile, it is viable to assume
that the quantum interference mechanisms that take place in the boundary of a
topological insulator also take place in this system. 
\begin{figure*}
\begin{center}
\includegraphics[width=0.3\textwidth,angle=0]{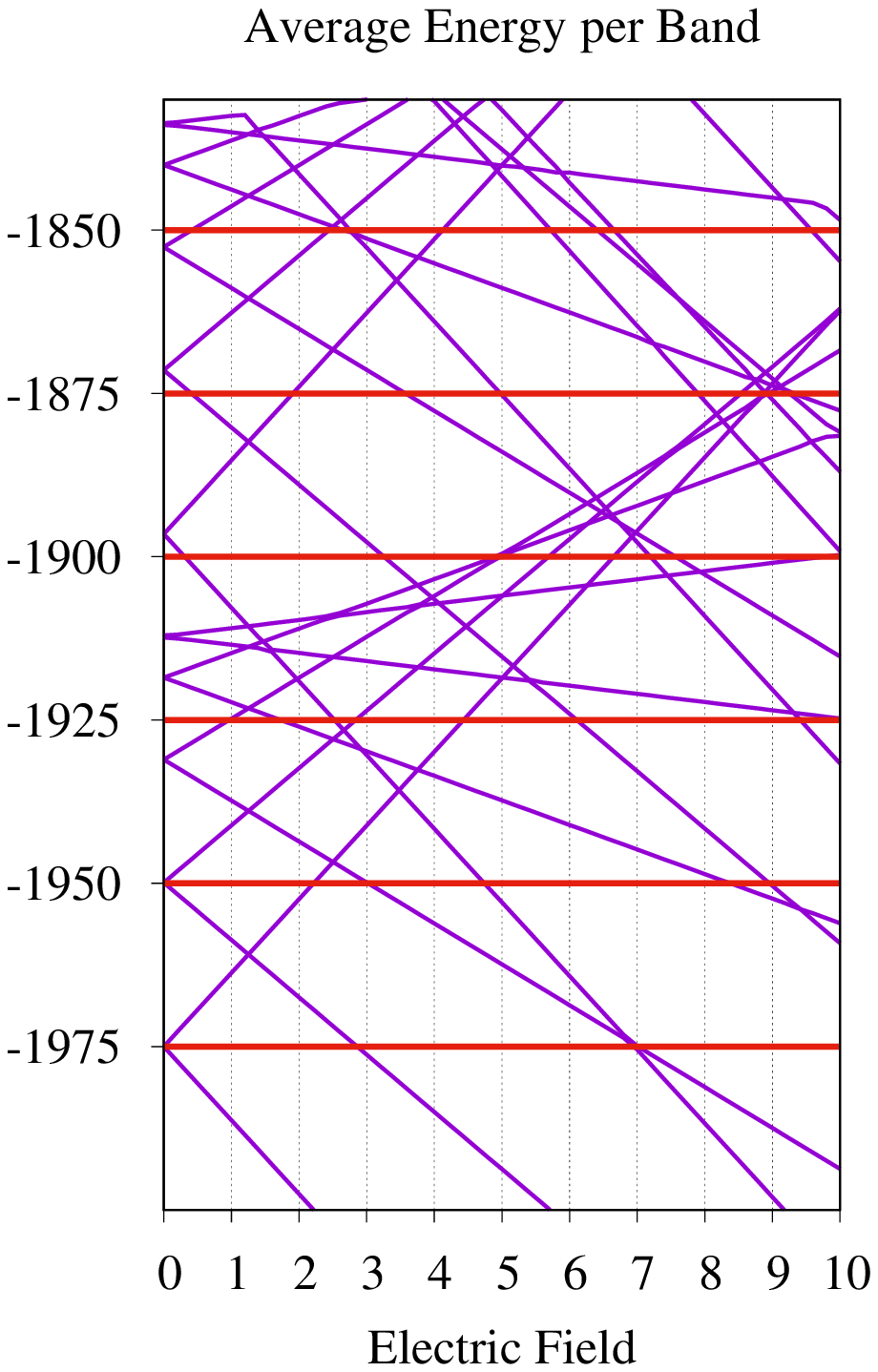}\includegraphics[width=0.6\textwidth,angle=0]{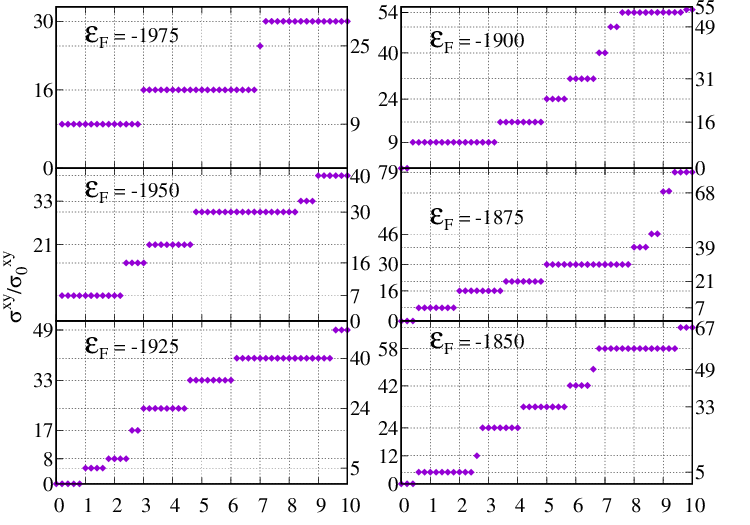} 
\caption{Left. Average value of energy over each band vs electric field, both in
atomic units. Red lines show the Fermi levels considered on the right. Right.
Zero-temperature transverse-conductivity $\sigma^{xy} / \sigma_0^{xy}$ vs Electric
Field (atomic units) for different values of the Fermi level. It can be seen that
conductivity always comes in integer multiples of $\sigma^{xy}_0$. Comparing with
the graph on the left it can be seen that the conductivity jumps every time a band
crosses the respective Fermi level. The system's parameters are indicated in the
caption of figure \ref{f06082}. For these parameters the reference value
$\sigma_0^{xy} = 13.78 \alpha$ was found analysing the conductivity data. The cases
$\mathscr{E}_F = -1975$ and $\mathscr{E}_F = -1950$ are distinctive in that in the
limit of zero electric field conductivity is nonvanishing.} 
\label{f06101}
\end{center}
\end{figure*}
\section{Conductivity}
The mean values of momentum and position over state (\ref{mar27})
are
\begin{gather}
\langle \hat{p}_x  \rangle_k = k + \sum_{j,q} |c_{j,q}(k)|^2 \frac{2 \pi j}{L} 
\end{gather}
Taking the effect of electron-electron collisions as a perturbation, it can be said
that at zero temperature the contribution of a given band to the $x$-conductivity
is at first order proportional to the sum of momentum mean values over that
band
\begin{gather}
\Pi_{x}^{band} = \sum_{l=-\frac{N}{2}}^{\frac{N}{2}} \langle \hat{p}_x \rangle_{k=\frac{2 \pi l}{L}}.   
\label{e07111}
\end{gather}
The transverse conductivity is proportional to the sum of contributions from all the
bands below the Fermi level $\mathscr{E}_f$
\begin{gather}
\sigma^{xy} = \alpha \sum_{band < \mathscr{E}_f} \Pi_{x}^{band}. 
\end{gather}
Being $\alpha$ a proportionality constant. Figure \ref{f06101} shows energy as well
as conductivity against electric field. It can be seen that $\sigma^{xy}$ comes in
integer multiples of a constant that shows no dependence with the electric field.
This conductivity displays a stair pattern, being constant over intervals of
different extension and increasing by integer steps of different size in the same
points where a given band crosses the corresponding Fermi energy as the electric
field grows. The stair pattern became less appreciable in simulations with smaller
$\lambda$. In the quantum Hall effect, the integer that determines the conductivity
is given by the number of times the wave-function phase winds around the boundary
of a two-dimensional Brillouin zone \cite{thouless,kohmoto}. Such an integer is
known in topology as the Chern invariant \cite{kane}. This parallel does not apply
here since the inclusion of the electric field in the Hamiltonian breaks
translational invariance in the $y$-axis and the reciprocal lattice becomes
one-dimensional. Whether there are additional topology constructs that apply in
this context or the system's discreteness can be ascribed to deeper precursors
remains to be seen. Whatever the case, this result shows that the conductivity's
quantization does not depend on the linear-response assumptions necessary to obtain
the Kubo formula \cite{kubo,thouless}. Another curious trait of figure \ref{f06101}
is that for the lowest two values of Fermi energy conductivity features finite
limits at zero electric field. For this to happen the Fermi energy must equal one
of the system's energy values for vanishing electric field. This limit is
consistent with a superconductor state because it features dissipationless
conduction on account of the quantum interference effect already discussed in this
document. It would then be interesting to add a magnetic field to Hamiltonian
(\ref{mahe}) and see whether this magnetic field is offset by the induced
magnetization, giving in this way a solid evidence of the superconducting state. 

Contrary to transversal conductivity, the first order longitudinal conductivity
vanishes for any electric field. This happens because the system displays
backscattering channels in the $y$-axis,
\begin{gather}
\hat{H}_0 | x,p_y,s_z \rangle = \hat{H}_0 | -x,-p_y,s_z \rangle.
\label{e07061}
\end{gather}
These channels are nonetheless displaced, so that TR symmetry is broken locally
just as in the Hall effect. By a similar mechanism, longitudinal conductivity could
be induced adding an electric field in the $x$-axis. Likewise, the inclusion of
interaction terms directly in the Hamiltonian would open a gap between the states
involved in (\ref{e07061}), since dissipation channels are spatially separated and
a particle would normally experience collisions in going from $x$ to $-x$ in
proportion to the magnitude of $x$. The same phenomenon can also explain the
longitudinal conduction in topological insulators \cite{konig}.  Although
electron-electron interaction was not considered in this work, the single body
functions obtained in section \ref{l0915} are the starting point to build a
second-quantization Hamiltonian giving a more accurate representation of the
system, in which case it is reasonable to expect nonvanishing
longitudinal-conduction. 
\section{Conclusions}
A single-body spin-orbit interaction model has been used to study the
conductivity pattern produced by a symmetry-breaking mechanism that takes place
in the system's bulk.  The electric field responsible for charge transport has
been included in the Hamiltonian and the study has been carried out by direct
diagonalization in order to explore the system's response beyond the linear
approximation.  Such a response displays a discrete pattern that is consistent
with the quantization of conductivity over the range of fields considered in
the study, showing in this way that a quantized conductivity does no depend on
linearity approximations. The quantization value is found to depend on the number
of energy bands located below the system's Fermi energy and also on the
intensity of the electric field.  In the particular case of the Fermi energy
being equal to an eigenenergy of the zero-field Hamiltonian, the limit of
conductivity for vanishing electric fields proves to be finite.  This feature
is consistent with a superconducting state but additional tests are necessary
to confirm such a hypothesis. Were it verified, it would mean the mechanism by
which the TR symmetry is broken in a superconductor can be related to the one
taking place on the edges of topological insulators. Overall, both the system's
physics as well as the perspective granted by the numerical analysis display
interesting features.
\section{Acknowledgments}
Financial support from Vicerrector\'ia de Investigaciones, Extensi\'on y
Proyecci\'on Social from Universidad del Atl\'antico through the program
Investigaci\'on Formativa is gratefully thanked by both authors.
\end{document}